\documentclass[twocolumn,superscriptaddress]{revtex4-2}

\usepackage{amssymb}
\usepackage{bm,amsmath}
\usepackage{graphicx} 
\usepackage{array}
\usepackage{setspace}
\usepackage{xcolor}
\usepackage{hyperref}
\usepackage{gensymb}
\usepackage{float}
\usepackage{soul}
\usepackage{ulem}
\hypersetup{
	colorlinks,
	linkcolor={red!90!black},
	citecolor={black!10!blue},
	urlcolor={blue!80!black}
}

\usepackage{newtxtext,newtxmath}

\newcommand{\f}{\frac}

\renewcommand{\r}{\textcolor{black}}

\newcommand{\Dreff}{D_r^\text{eff}}

\newcommand{\D}{D^\text{eff}}
\newcommand{\cpn}{\chi_4^\text{peak}}
\newcommand{\tpeak}{t_\text{peak}}

\begin{document}
\title{\r{Growing length and time scales in} activity-mediated glassy dynamics in confluent cell monolayers}
	\author{Souvik Sadhukhan}
	\email{ssadhukhan@tifrh.res.in}
	\affiliation{Tata Institute of Fundamental Research, 36/P Gopanpally Village, Hyderabad - 500046, India}
	\author{Chandan Dasgupta}
	\email{cdgupta@iisc.ac.in}
	\affiliation{Department of Physics, Indian Institute of Science, Bangalore 560012, India}
	\affiliation{International Centre for Theoretical Sciences, TIFR, Bangalore 560089, India}
    \author{Saroj Kumar Nandi}
    \email{saroj@tifrh.res.in}
    \affiliation{Tata Institute of Fundamental Research, 36/P Gopanpally Village, Hyderabad - 500046, India}

\begin{abstract}
Activity-mediated unjamming of a confluent glassy system is crucial for several \r{biological processes, such as embryogenesis and cancer metastasis}. During these processes, the cells progressively change their junction properties, characterized by an interaction parameter $p_0$, and become motile. \r{Here, we study the effect of nonequilibrium active fluctuations, in the form of self-propulsion, on the glassy dynamics in a confluent system.} We simulate the active Vertex model and use the analytical mode-coupling theory (MCT) to show that the nature of the transition in the presence of activity remains similar to that in \r{a thermal system where the fluctuations are temperature-like}. The agreement of the simulation results with the MCT predictions demonstrates that the structure-dynamics feedback mechanism controls the relaxation dynamics. In addition,  we present the first computation of a dynamic length scale, $\xi_d$, in confluent systems  \r {using finite-size scaling,} and show that the growing relaxation time \r{exhibita a power-law dependence on} $\xi_d$. Furthermore, unlike particulate glasses, the static length \r{that governs the finite-size scaling of the relaxation time} is proportional to $\xi_d$, revealing the unique nature of the glassy dynamics in confluent systems.

\end{abstract}
\maketitle

\section{Introduction}
\r{Many past works have shown that an epithelial monolayer of cells shows glassy behavior where the system has anomalously slow dynamics with a stretched exponential relaxation \cite{park2015,activereview}, non-Gaussian displacement distribution \cite{sadati2013,park2016,activereview}, spatially heterogeneous dynamics \cite{angelini2011,park2015}, etc. The glassy dynamics and unjamming of epithelial layers are relevant to many biological processes.}
Metastasis is the leading cause of death in cancer patients \cite{steeg2006,fares2020}. It has several steps, where the cancer cells leave their primary sites, go through the stroma and other tissues, use the bloodstream, and acclimatize at a secondary site \cite{wirtz2011,lee2019,massague2016,steeg2006}. The primary step of this process is cellular unjamming \cite{oswald2017,mitchel2020}.  The normal epithelial monolayer is primarily sedentary, where cells show strong cell-cell adhesion and out-of-plane polarity. By contrast, the cancerous monolayer is dynamic, where cells show weak cell-cell adhesion, develop in-plane polarity, and become motile \cite{mitchel2020,thiery2002,yang2020}.
During several biological processes, such as embryogenesis \cite{mongera2018,frieder2002,hannezo2014} and cancer progression \cite{friedl2009,kumar2009,hanahan2011,malinverno2017,kim2020,mitchel2020}, the cells undergo an epithelial-to-mesenchymal transition (EMT) \cite{thiery2002,thery2006,hugo2007,lu2013}.  Although EMT was thought of as a binary switch between the epithelial and mesenchymal states, it is now clear that it is more like a continuum than a switch \cite{lu2013,tian2013,zhang2014,mohit2015}. At the early stage of EMT, known as partial EMT or pEMT, the monolayer remains confluent, although the cells become motile. Cellular motility facilitates unjamming \cite{mitchel2020,bi2016}; a more detailed characterization of the transition is essential for a deeper understanding of metastasis \r{and other biological processes mentioned above}.

Experiments have shown contrasting results for different oncogenes. For example, human breast cancer cells MCF-10A fluidizes by the over-expression of an oncogene, $14-3-3\zeta$, or an endocytic protein, RAB5A \cite{malinverno2017}. However, the same monolayer solidifies by over-expressing another oncogene ErbB2/HER-2/neu \cite{sadati2013}. A recent work has shown that confluency has a nontrivial effect on activity, where the former leads to an effective rotational diffusivity, $\Dreff$, that is different from the intrinsic rotational diffusivity, $D_r$, of motility \cite{sadhukhan2024}. Thus, how activity will affect the glassy dynamics is nontrivial. Furthermore, simulations suggest that within the continuum of pEMT \cite{sadhukhan2024,li2021}, during the initial times when cellular junctions are relatively strong, the monolayer shows super-Arrhenius relaxation. Conversely, at a later time, when junctions become weaker, the monolayer shows sub-Arrhenius relaxation. In this work, we focus on this later regime. Using large-scale simulations of the active Vertex model and an analytical theory, the mode-coupling theory (MCT) of glasses \cite{gotze1992,das2004,janssen2018}, we investigate the effects of motility on the unjamming transition. What are the natures of the glassy dynamics and this transition? How does the dynamics of a monolayer with pEMT differ from that of the epithelial states? 
\r{In a recent work, some of us have shown that the sub-Arrhenius regime of the dynamics in the thermal Vertex model agrees well with the predictions of a particular theory, the mode-coupling theory (MCT) \cite{pandey2023}, of equilibrium glasses.} MCT advocates a feedback mechanism where the static structure strongly correlates with the glassy dynamics, leading to a dramatic slowing down. \r{Reference \cite{pandey2023} also showed that the theoretical predictions qualitatively agree with experiments on confluent cell monolayers. In this work, we explore if the structure-dynamics correlation of confluent systems \cite{atia2018,sadhukhan2022,pandey2023} survives even under motility and if the nature of the unjamming transition remains similar.}

\r{Activity can have nontrivial effects on glassy systems, leading to reentrant dynamics, tunable fragility, etc \cite{Berthier2017,Debets2021,puneet2025}. Furthermore, recent simulations have shown that the dynamic heterogeneity (DH) in active glasses can have remarkably different behavior compared to passive systems \cite{paul2023,ghoshal2020,paoluzzi2021}; for example, the DH length scale can be enormous and can grow even when the relaxation time, $\tau$, remains the same \cite{paul2023}. The active inhomogeneous MCT shows that activity leads to an additional source term governing the behavior of DH \cite{kolya2024}. However, the nature of confluent systems is different compared to particulate systems. It is unclear how activity will affect the glassy dynamics in these systems. Thus, a quantitative characterization of  activity-mediated unjamming of confluent glassy systems and the role of the DH in the dynamics is indispensable. We take up this task in this work; here are our main results. We show that the relaxation time, $\tau$, diverges at a critical self-propulsion velocity, $v_c^2$. The data in our simulation range agrees well with the MCT predictions. In addition, we have also studied the behavior of the four-point correlation function and present the first computation of the dynamic length scale, $\xi_d$, in confluent systems \r{from a finite-size scaling analysis of the four-point susceptibility}. We find that $\xi_d$ also diverges at the same critical point, $v_c^2$ and $\xi_d \sim \tau^{1/3}$. We have also computed a static length scale, $\xi_s$ \r{from finite-size scaling of the relaxation time}. We find that $\xi_s\propto \xi_d$: this contrasts the behavior in particulate systems and reveals the unique nature of the glassy dynamics in confluent systems. We organize the manuscript as follows: We provide the details of the model in Sec. \ref{model}. We present the results for the two-point correlation function and the relaxation dynamics in Sec. \ref{resultstwopoint} and those for the four-point correlation function and the length scales in Sec. \ref{resultsfourpoint}. We conclude the paper by discussing our results in Sec. \ref{discussion}.
}

\section{The model}
\label{model}
\r{We have simulated an athermal active Vertex model \cite{bi2016,sadhukhan2024} of confluent epithelial monolayers. This model represents the cells as polygons \cite{Marder1987,Weaire2001,farhadifar2007,Barton2017,albert2016} with an area and a perimeter constraint. The energy function, $\mathcal{E}$, governing the dynamics of the model is 
\begin{equation}\label{hamiltonian}
	\mathcal{E} = \sum_{i=1}^{N} \Big[ \Lambda_A (A_i - A_0)^2 + \Lambda_P (P_i - P_0)^2 \Big ],
\end{equation} 
where $N$ is the total number of cells, $A_0$ and $P_0$ are the target area and target perimeter, respectively. $A_i$ and $P_i$ are the area and perimeter of the $i$th cell and $\Lambda_A$ and $\Lambda_P$ are area and perimeter moduli. The area term in Eq. (\ref{hamiltonian}) comes from the incompressibility property \cite{jacques2015} of cell cytoplasm. On the other hand, $P_0$ encodes the cortical properties and cell-cell adhesion  \cite{farhadifar2007,Barton2017,jacques2015}. We take the unit of length as $\sqrt{A_0}$ and rescale the energy function, Eq. (\ref{hamiltonian}), by $A_0$ to obtain 
\begin{equation}\label{energyfunction}
	\mathcal{H}=\frac{\mathcal{E}}{A_0^2}=\sum_{i=1}^N\bigg[ \lambda_A(a_i-1)^2+\lambda_P(p_i-p_0)^2\bigg],
\end{equation}
where $\lambda_A=\Lambda_A$, $\lambda_P=\Lambda_P/A_0$, $a_i=A_i/A_0$, $p_i=P_i/\sqrt{A_0}$, and $p_0=P_0/\sqrt{A_0}$. Using the energy function $\mathcal{H}$, we can calculate the force on the vertices,  $\mathbf{F}_j=-\nabla_j\mathcal{H}$. In addition, there is also an active force, $\mathbf{f}_a^j$. We integrate the over-damped equations of motion for each vertex $j$ given by,
\begin{equation}\label{eqofmotion}
	\frac{d{\bf{r}}_j}{dt} = \mu^{-1} \Big({\bf{F}}_j + {\bf{\tilde{f}}}_a^{j} \Big),
\end{equation}
where $\mu$ is the friction coefficient, set to $1$.}

\r{For self-propulsion, we assign a polarity vector with each cell: $\hat{\mathbf{n}}_i=(\cos\theta_i,\sin\theta_i)$, where $\theta_i$ is the angle with the $x$-axis. We obtain the active force on the vertex $j$ as $\mathbf{f}_a^j=f_0\hat{\mathbf{n}}_j=\mu v_0\hat{\mathbf{n}}_j$. We compute $\mathbf{f}_a^j$ as ${\bf \tilde{f}}_a^j = \frac{v_0}{3} \sum_{l \in \mathcal{N}(j)} \hat{\mathbf{n}}_l$, where $\mathcal{N}(j)$ is the number of all neighbouring cells sharing vertex $j$. $\theta_i$ performs rotational diffusion and is governed by the equation \cite{bi2016}, 
\begin{equation}\label{thetaeq}
	\partial_t \theta_i(t) =\sqrt{2D_r} \eta_i(t)
\end{equation}
where $\eta_i$ is a Gaussian white noise, with zero mean and a correlation $\langle \eta_i(t)\eta_l(t^\prime)\rangle = \delta(t-t^\prime)\delta_{il}$ and we have set the rotational friction coefficient to unity. $D_r$ is the rotational diffusion coefficient that is related to a persistence time, $\tau_p=1/D_r$.
We have taken a $50:50$ (number ratio) binary mixture with $a_{0\alpha} = 0.8$ and $a_{0\beta} = 1.2$ to avoid the periodic hexagonal ground state, and designate the system via $p_0=p_{0\alpha}/\sqrt{a_{0\alpha}}=p_{0\beta}/\sqrt{a_{0\beta}}$. 
We present the results using $1/\lambda_A\mu a_0$ as the unit of time.}

\begin{figure*}
	\centering
	\includegraphics[width=16.5cm]{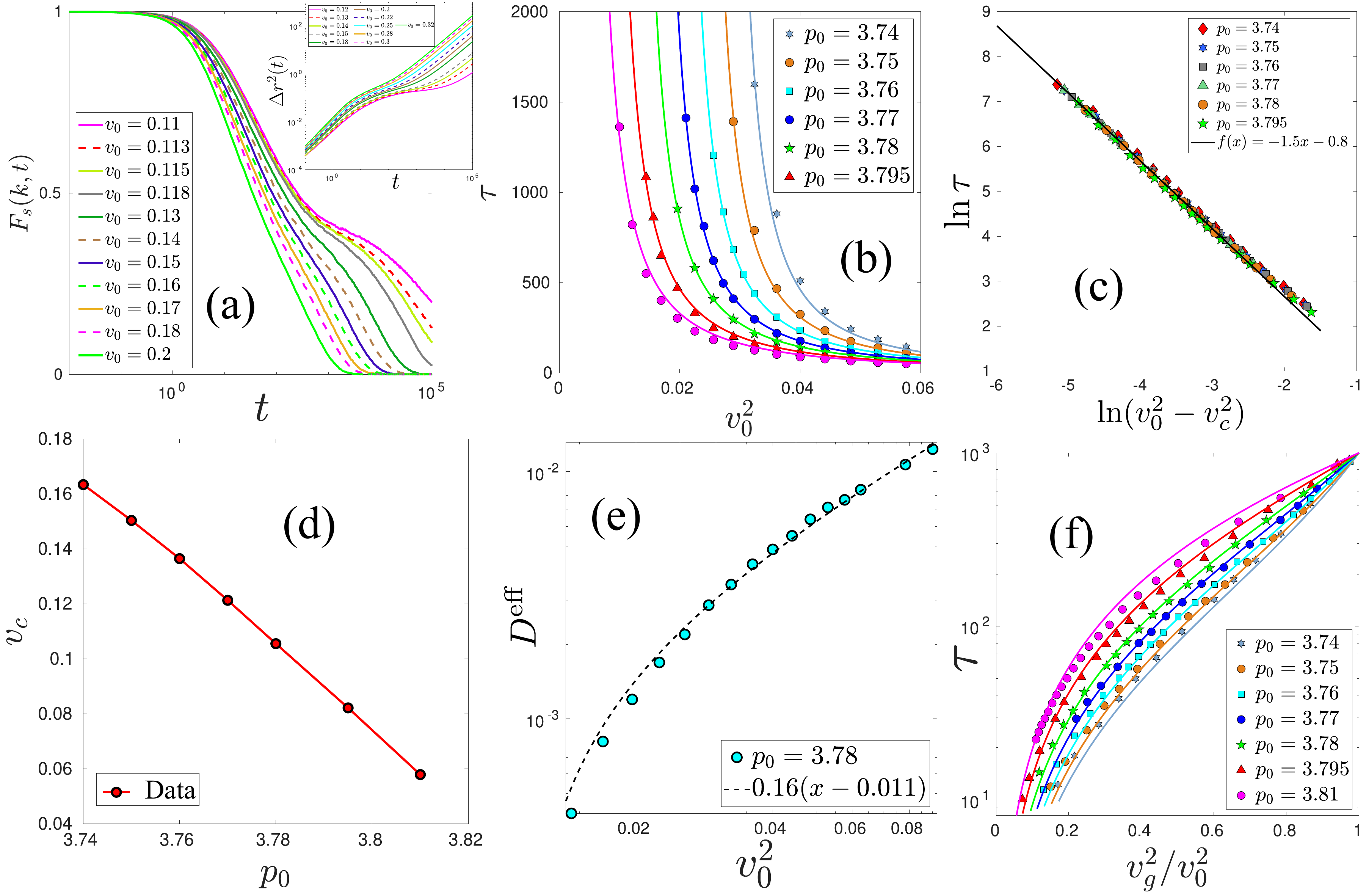}
	\caption{Comparison of simulation data on \r{the active Vertex model} with MCT: (a) Evolution of the self intermediate scattering function, $F_s(k,t)$ for various $v_0$ and $p_0=3.78$. {\bf Inset:} Mean-square displacement for the corresponding parameters as in the main figure. (b) The relaxation time, $\tau$, as a function of $v_0^2$ for different $p_0$. The points are simulation data, and the lines denote fits with Eq. (\ref{MCTrelaxationPowerlaw}). (c) Plot of $\log(\tau)$ as a function of $\log(v_0^2  - v_c^2)$ for various $p_0$ shows a data collapse to a master curve as $\gamma=3/2$ and $A$ are constants. The line represents a linear fit with the slope $3/2$. (d) The MCT transition point, $v_c$, monotonically decreases with $p_0$. (e) The effective diffusion constant, $\D$, goes to zero as a power-law as $v_0$ tends to $v_c$, but with a different exponent than $\gamma$, signifying the breakdown of the Stokes-Einstein relation. (f) Angell plot representation of $\tau$: $\log(\tau)$ is plotted against $v_g^2/v_0^2$. $v_g$ is defined as the value of $v_0$ when $\tau$ reaches $10^3$. It shows a sub-Arrhenius relaxation. Data presented here are for $D_r = 1$ and $N = 256$.}
	\label{mctcompare}
\end{figure*}

To take advantage of the effective equilibrium scenario at small $\tau_p$, we fix $D_r=1/\tau_p=1$ and study the properties with varying $f_0$ \cite{parisi2005,nandi2017,Flenner2016}. Since we set $\mu=1$, $f_0$ is numerically the same as the self-propulsion velocity, $v_0$. We present the results in terms of $v_0$ using $v_0^2$ in place of temperature $T$ to characterize the transition. Reference \cite{pandey2023} has shown that MCT works surprisingly well for epithelial systems. Here, we take a similar approach. We test the applicability of the MCT by comparing the relaxation dynamics with the MCT predictions. We show that MCT remains valid even in the presence of activity; thus, the structure-dynamics correlation holds, and the feedback mechanism of MCT controls the dynamics. The relaxation time, $\tau$, diverges with the same exponent, $\gamma=3/2$, as in equilibrium \cite{pandey2023}; this indicates that the relaxation dynamics remains equilibrium-like, much like that in particulate active systems at small $\tau_p$ \cite{activereview,nandi2017,Berthier2019c,Flenner2016,paul2023}. In addition, we present the first computations of the dynamic length scale, $\xi_d$, in confluent systems. $\xi_d$ also diverges at the critical point, and the increasing $\tau$ accompanies a growing $\xi_d$.

\section{Results}

\subsection{Two-point correlation function and the relaxation dynamics}
\label{resultstwopoint}
We characterize the dynamics via the self-intermediate scattering function, $F_s(k,t)$, defined as
\begin{equation}\label{fskteq}
	F_s(k,t) = \langle \tilde{F}_s(k,t) \rangle =\frac{1}{N_\alpha} \langle \sum_{i=1}^{N_\alpha} \exp[i\mathbf{k}.(\mathbf{r}_i(0) - \mathbf{r}_i(t))]\rangle
\end{equation}
where, $k$ is the magnitude of the wave-vector $\mathbf{k}$, $N_\alpha$, the number of cells with target area $a_{0\alpha}$, $\mathbf{r}_i$, the centre of mass of the $i^{\text{th}}$ cell. \r{The angular brackets denote averages over both ensembles and different time origins}. We consider only the $\alpha$-particles for the calculation of $F_s(k,t)$ \cite{szamel2006,pareek2023}. We present the results \r{for $k=k_\text{max}=4.5$} corresponding to the first peak of the structure factor. Figure \ref{mctcompare}(a) shows the decay of $F_s(k,t)$ for various values of $v_0$. The inset of Fig. \ref{mctcompare}(a) shows the mean square displacement, $\Delta r^2(t)$, defined as
\begin{align}
	\Delta r^2(t)=\left\langle{\f{1}{N}\sum_{i=1}^N (\mathbf{r}_i(t)-\mathbf{r}_i(0))^2} \right\rangle.
\end{align}
The characteristic two-step decay of $F_s(k,t)$ is evident at lower values of $v_0$. Similarly, $\Delta r^2(t)$ also shows a sub-diffusive behavior at intermediate times and becomes diffusive at long times (inset of Fig. \ref{mctcompare}a).

We define the relaxation time, $\tau$, as $F_s(k,\tau)=0.3$. \r{Reference \cite{pandey2023} has shown that MCT works well in the sub-Arrhenius relaxation regime in a confluent cellular system, where the fluctuations are thermal. MCT is a critical theory of equilibrium glasses that relies on a structure-dynamics feedback mechanism \cite{gotzebook}. The static structure gets affected as temperature, $T$, decreases; this slows down the dynamics, which again affects the static properties. This mechanism leads to a critical theory for glassy dynamics where $\tau$ diverges at the critical point, $T_c$. Much like ordinary critical pheonomena, MCT also predicts a power-law divergence of $\tau$,
\begin{equation}
		\tau = \tilde{A}(T-T_c)^{-\gamma},
\end{equation}
where $\gamma$ is a universal exponent. As we are dealing with an active athermal system at constant $\tau_p$, $v_0^2$ plays the role of an effective temperature, and we can use $T \sim v_0^2$ in the small $\tau_p$-regime. Therefore,} translating this result for active systems at small $\tau_p$, we obtain 
\begin{equation}\label{MCTrelaxationPowerlaw}
\tau = A(v_0^2 - v_c^2)^{-\gamma},
\end{equation}
where $v_c$ is the MCT critical point, $A$ is a constant, and $\gamma$ is an exponent. We obtain $\tau$ in our simulation with varying $v_0$ for several values of $p_0$. For a particular $p_0$, we fit the data of $\tau$ as a function of $v_0$ with Eq. (\ref{MCTrelaxationPowerlaw}) and obtain $A$, $\gamma$, and $v_c$. We show these fits for several values of $p_0$ in Fig. \ref{mctcompare}(b). We find that $A=0.71$ and $\gamma=3/2$ remain constant for various $p_0$ in the regime of our interest here. Figure \ref{mctcompare}(c) shows the plot of $\ln\tau$ as a function $\ln(v_0^2-v_c^2)$: Eq. (\ref{MCTrelaxationPowerlaw}) predicts that this should be a straight line with slope $-3/2$ (solid line). The simulation data agree remarkably well with the MCT prediction, Eq. (\ref{MCTrelaxationPowerlaw}). \r{We note that this value is slightly smaller than the lower limit of $\gamma$ ($\simeq 1.76$) that MCT predicts for equilibrium glassy systems. The reason is possibly the nonequilibrium nature of the system \cite{pandey2023,lama2022}}. Much like any critical theory, the critical point depends on the system parameters. In the context of the vertex model, $p_0$ parametrizes the interaction potential; therefore, we expect $v_c$ to vary with changing $p_0$. Figure \ref{mctcompare}(d) shows that $v_c$ monotonically decreases as $p_0$ increases. We emphasize that although $v_c$ changes with $p_0$, the exponent $\gamma$ in Eq. (\ref{MCTrelaxationPowerlaw}) remains constant; this is consistent with the universal prediction of the theory. \r{We emphasize that this is in contrast to particulate systems, where the values of $\gamma$ are different for various systems.}

MCT also predicts a power-law decay of $F_s(k,t)$ around the plateau ($\beta$-regime), and the simulation results are consistent with these predictions (see Appendix \ref{mctdecayexpo}).

We have also computed the effective diffusion constant, $\D$, as the ratio of self-diffusivity and free diffusion constant of an isolated cell. $\D = D_s/D_0$ where $D_s = \text{lim}_{t\rightarrow\infty} \langle \Delta r^2(t) \rangle/(4t)$ and $D_0 = v_0^2/2D_r$. $\D\to0$ when $\tau$ diverges. Fitting the data of $\D$ with the power law form of MCT, $\D=A'(v_0^2-v_c^2)^{\gamma'}$, we obtain the same $v_c$ as from the data of $\tau$, and $\gamma'=1$. We show the fit of $\D$ for $p_0=3.78$ in Fig. \ref{mctcompare}(e). Note that the exponent $\gamma$ in Eq. (\ref{MCTrelaxationPowerlaw}) and $\gamma'$ are different, implying the violation of the Stokes-Einstein relation \cite{einsteinpaper,cicerone1996,pareek2023}. \r{This breakdown is similar to what one finds in particulate systems, where MCT predicts that the relation remains valid, but simulations show a violation \cite{flenner2005} (see however Ref. \cite{anshell2024})}. We will explore this breakdown of the Stokes-Einstein relation in detail in a separate work. Finally, we show the Angell plot representation of $\tau$ as a function of $v_g^2/v_0^2$, where we have defined $\tau(v_g)=10^3$ in Fig. \ref{mctcompare}(f). Consistent with the agreement of simulation results with MCT, the relaxation dynamics shows a sub-Arrhenius behavior \cite{pandey2023}.

\begin{figure}
	\centering
	\includegraphics[width=8.4cm]{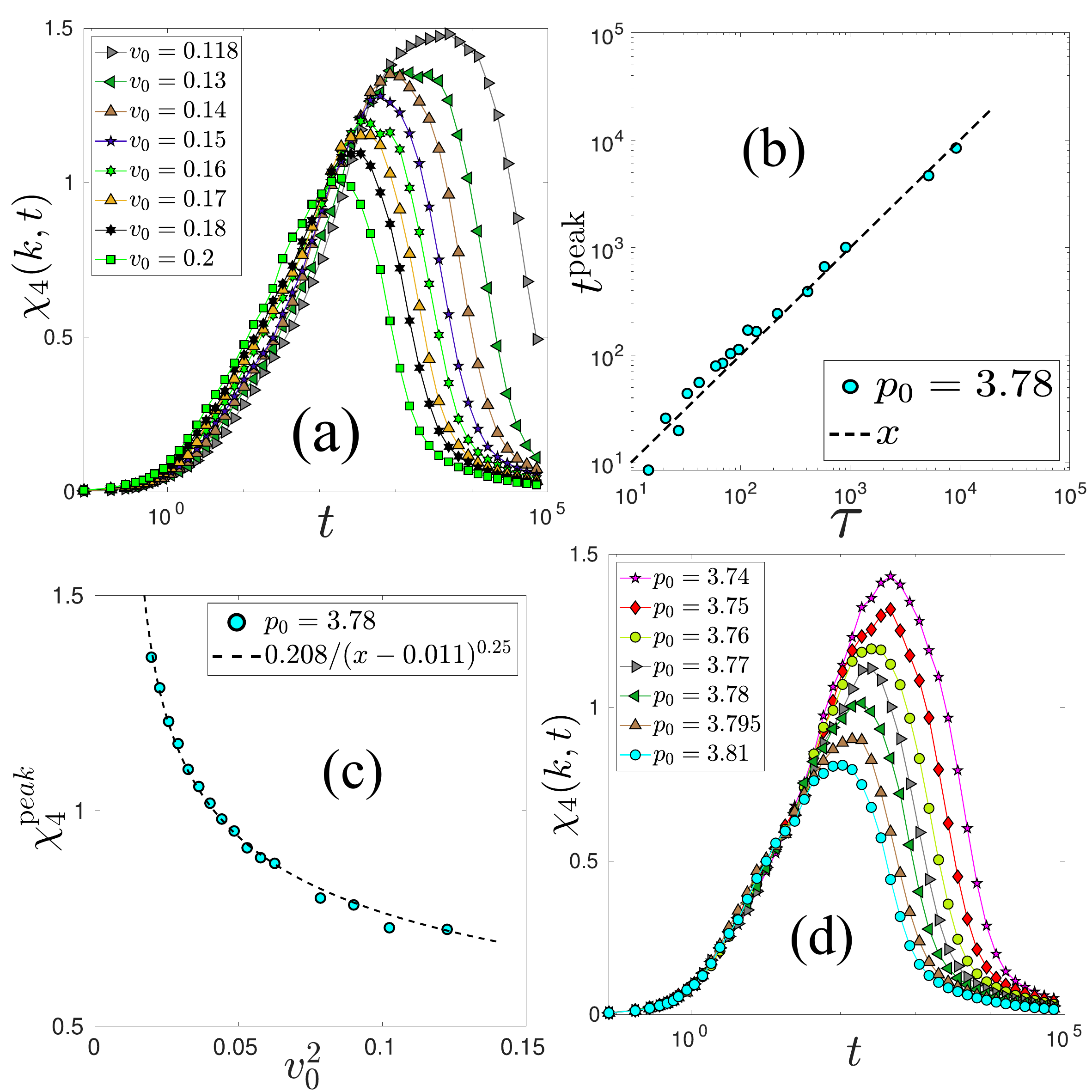}
	\caption{The behavior of the four-point correlation function, $\chi_4(k,t)$. (a) $\chi_4(k,t)$ has a non-monotonic dependence on $t$. It grows at short times, reaches a peak, $\cpn$, at time $t_\text{peak}$, and then decays towards zero. (b) $t_\text{peak}$ gives another measure of the relaxation time and is proportional to $\tau$. The dashed line is a linear fit of the data (symbols). (c) $\cpn$ diverges as a power law at $v_c$ with the exponent $1/4$. (d) Evolution of $\chi_4(k,t)$ at $v_0=0.2$ with varying $p_0$. Both $\cpn$ and $t_\text{peak}$ increase as $p_0$ decreases.}
	\label{chi4compare}
\end{figure}

\subsection{Four-point correlation function and the length scales}
\label{resultsfourpoint}
Next, we calculate the four-point susceptibility, $\chi_4(k,t)$, defined as the fluctuations in $F_s(k,t)$, as follows,
\begin{equation}\label{chi4eq}
	\chi_4(k,t) = N_\alpha[\langle \tilde{F}_s(k,t)^2\rangle -  \langle \tilde{F}_s(k,t) \rangle^2],
\end{equation}
where $\tilde{F}_s(k,t)$ is defined in Eq. (\ref{fskteq}). $\chi_4(k,t)$ gives the measure of dynamic heterogeneity (DH) in glassy systems. As $t$ increases, $\chi_4(k,t)$ grows from zero, reaches a maximum, and then decays to zero at long times. We show the behavior of $\chi_4(k,t)$ for $k$ corresponding to the structure factor maximum and $p_0=3.78$ for various $v_0$ in Fig. \ref{chi4compare}(a). As $v_0$ decreases, the peak height of $\chi_4(k,t)$, $\cpn$ increases. The time, $\tpeak$, at which $\chi_4(k,t)$ reaches its maximum, gives another measure of the relaxation time. Typically, one finds $\tpeak\propto \tau$ \cite{nandi2012,activereview}; Fig. \ref{chi4compare}(b) shows that this relation holds for the active Vertex model. 
Furthermore, $\cpn$ gives a measure of the correlation volume. Figure \ref{chi4compare}(a) shows that $\cpn$ increases as $v_0$ decreases. We compare $\cpn$ with the power-law prediction of MCT: $\cpn\sim (v_0^2-v_c^2)^{-{\delta}}$. The simulation data shows that $\cpn$ also diverges at the same $v_c$ with $\delta=1/4$. We show the fit for $p_0=3.78$ in Fig. \ref{chi4compare}(c); the behavior for other values of $p_0$ remains similar.

\begin{figure*}
	\centering
	\includegraphics[width=17cm]{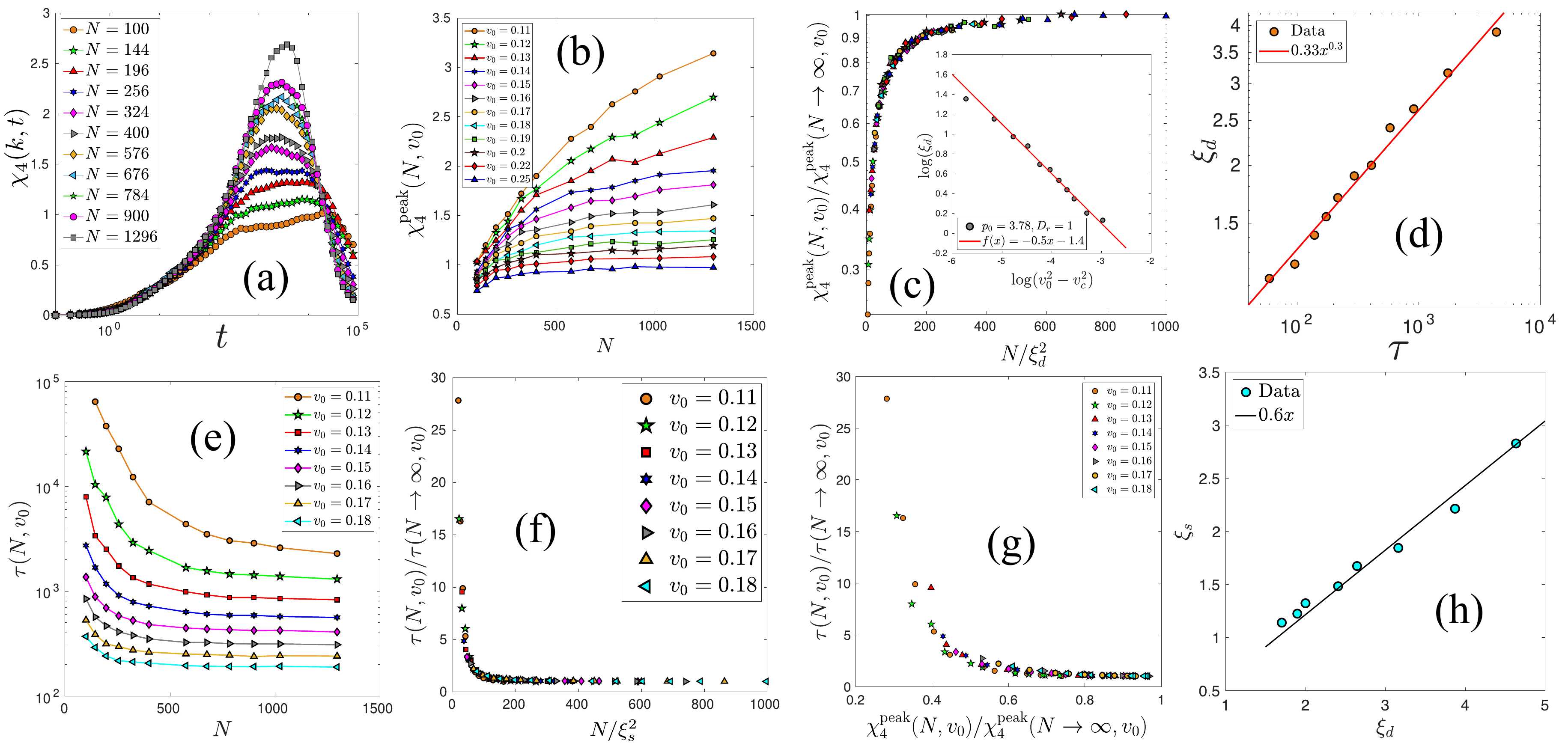}
	\caption{Finite-size scaling analysis of $\cpn$ and $\tau$. (a) $\chi_4(k,t)$ as a function of $t$ for different values of $N$ for $v_0 = 0.12$ and $p_0 = 3.78$. $\cpn$ increases and the peak becomes sharper as $N$ increases. (b)$\cpn(N,v_0)$ as a function $N$ for various $v_0$. $\cpn$ initially increases with $N$ and then saturates to a $v_0$-dependent value. (c) Scaling collapse of $\cpn(N,v_0)/\cpn(N\rightarrow\infty,v_0)$ as a function of $N/\xi_d^2(v_0)$. The data collapse for the specific values of $\xi_d(v_0)$ gives the dynamic length scale. {\bf Inset:} Plot of $\log(\xi_d)$ as a function of $\log(v_0^2 - v_c^2)$ shows a linear trend, signifying a power-law behavior. We find the exponent $\nu =1/2$ from a linear fit (line) of the data (symbols). (d) MCT predicts a power-law relation between $\xi_d$ and $\tau$: $\xi_d\propto\tau^{1/z}$ with $z=\gamma/\nu=3$. The simulation data (symbols) are consistent with this behavior (line). (e) $\tau$ as a function of $N$ for different $v_0$, showing that $\tau$ initially decreases with $N$ and then saturates to a $v_0$-dependent value for large $N$. (f) Scaling collapse of $\tau(N,v_0)/\tau(N\rightarrow\infty,v_0)$ as a function of $N/\xi_s^2(v_0)$ for appropriate choice of the length scales $\xi_s(v_0)$. (g) Plot of $\tau(N,v_0)/\tau(N\rightarrow\infty,v_0)$ versus $\chi_4^{\text{peak}}(N,v_0)/\chi_4^{\text{peak}}(N\rightarrow\infty,v_0)$ show data collapse for different $v_0$ and $N$. (h) Plot of $\xi_s(v_0)$ versus $\xi_d(v_0)$ show that they are proportional to each other.}
	\label{finitesizescaling}
\end{figure*}

How does $\chi_4(k,t)$ behave at constant activity but varying $p_0$? Figure \ref{chi4compare}(d) shows $\chi_4(k,t)$ at constant $v_0=0.2$ and different values of $p_0$. Park {\it et al.} \cite{park2015} showed that the monolayers of both asthmatic and non-asthmatic human bronchial epithelial cells become more sluggish as they mature with passing days. As the monolayer matures, the cell-cell junctions become more firm, leading to a decrease in $p_0$. In the experiments, it results in decreasing values of the observed perimeter or the shape index \cite{park2015}. Figure \ref{chi4compare}(d) shows that as $p_0$ decreases, $\tpeak$ and $\cpn$ increase. These results rationalize the experimental findings: As the system matures and the junctions become stronger, the system becomes more sluggish, and the volume of dynamically heterogeneous regions increases. We can also obtain the DH length scale, $\xi_d$, from the data of $\cpn$ via a finite size scaling, as we demonstrate below.

Within MCT, the glassy dynamics comes from a genuine phase transition where $\tau$ diverges concomitantly with the divergence of $\xi_d$. We first compute $\chi_4(k,t)$, Eq. (\ref{chi4eq}), for several system sizes with $N$ spanning from $N = 100-1296$, for different values of $v_0$ and $p_0 = 3.78$. We show $\chi_4(k,t)$ for $v_0 = 0.12$ for different values of $N$ in Fig. (\ref{finitesizescaling}a); $\cpn$ grows as $N$ increases. Figure \ref{finitesizescaling}(b) shows $\cpn(N,v_0)$, the peak value of $\chi_4(k,t)$ for specific values of $N$ and $v_0$, as a function of $N$ for different values of $v_0$. $\cpn(N,v_0)$ initially grows with increasing $N$ and saturates to a $v_0$-dependent value as $N\to\infty$. This behavior indicates the presence of a length scale, $\xi_d$; $\cpn(N,v_0)$ continues to grow with increasing $N$ when $\xi_d$ is larger than the system size and then saturates in the other limit.

According to the finite-size scaling hypothesis, $\cpn(N,v_0)/\cpn(N\rightarrow\infty,v_0)$ should be a function of $N/\xi_d^2(v_0)$. To extract the correlation length, we find the values of $\xi_d(v_0)$ for which plots of
$\cpn(N,v_0)/\cpn(N\rightarrow\infty,v_0)$ as a function of $N/\xi_d^2(v_0)$for all $N$ and $v_0$ collapse onto a master curve. Note that in this method, the correlation length $\xi_d(v_0)$ is determined modulo a multiplicative constant: if data collapse is obtained for a set of $\xi_d(v_0)$, a similar data collapse would be obtained for $c\xi_d(v_0)$ where $c$ is a constant. The data collapse for $p_0=3.78$ is shown in Fig. \ref{finitesizescaling}(c), and the corresponding values of $\xi_d(v_0)$ are shown in the inset of Fig. \ref{finitesizescaling}(c). Following the scaling prediction of inhomogeneous mode-coupling theory (IMCT) for glassy systems \cite{Biroli2006} and using the effective equilibrium nature of our system, we obtain for our active Vertex model, $\xi_d(v_0) \sim (v_0^2-v_c^2)^{-\nu}$.
We have fitted $\xi_d$ as a function of $v_0^2$ with this power-law form and obtained the same $v_c$ as before, and $\nu=1/2$. The inset of Figure \ref{finitesizescaling}(c) shows the plot of $\log(\xi_d)$ as a function of $\log(v_0^2 - v_c^2)$ that follows a straight with the slope $\nu=1/2$. The scaling predictions of ordinary critical phenomenon suggest $\delta=\nu(2-\eta)$ \cite{kardarbook}. Using the values of $\delta$ and $\nu$, we obtain $2-\eta=1/2$. This value is quite different from the prediction of IMCT \cite{Biroli2006} and simulation results for passive particulate systems \cite{karmakar2009,karmakar2010}. However, the value is reasonable for a two-dimensional system where the two-point spatial order-parameter correlation function decays as $1/r^\eta$, where $r$ is the spatial distance. MCT is a critical theory that advocates a diverging correlation length, $\xi_d$, accompanying the divergence of $\tau$. Using the scaling relations for the individual variables, we obtain $\xi_d\sim \tau^{1/z}$ where $z=\gamma/\nu=3$. We show the behavior of $\xi_d$ as a function of $\tau$ in Fig. \ref{finitesizescaling}(d) and find the exponent $z\simeq 3$.

We have also checked the system-size scaling of the relaxation time, $\tau$, for the active confluent cell monolayer. \r{We can extract another length scale, the static correlation length, $\xi_s$, from the system-size dependence of $\tau$ \cite{karmakar2009}. Generally, $\xi_s$ and $\xi_d$ are distinct for passive particulate systems \cite{karmakar2009}. Given the distinctive nature of the confluent systems, we now check whether a similar result also holds for the active Vertex model.
Figure \ref{finitesizescaling}(e) shows $\tau$ vs. $N$ for various $v_0$}. As $N$ increases, $\tau$ decreases and eventually saturates to a $v_0$-dependent value. The saturation value increases with decreasing $v_0$. The system-size dependence of $\tau$ is similar to that in passive particulate systems~\cite{karmakar2009}. However, it contrasts with the usual dynamical finite-size scaling close to criticality in which the relaxation time increases with system size. We find that $\tau(N,v_0)$ also exhibits a scaling collapse when we plot $\tau(N,v_0)/\tau(N\rightarrow\infty,v_0)$ as a function of $N/\xi_s^2(v_0)$, Fig.~(\ref{finitesizescaling}f), for appropriate choice of $\{\xi_s(v_0)\}$. As shown in Fig.~(\ref{finitesizescaling}g), $\xi_s$ turns out to be proportional to $\xi_d$, establishing that the system-size dependences of $\cpn$ and $\tau$ are governed by the {\it same} correlation length (modulo a multiplicative constant). This is analogous to finite-size scaling in usual critical phenomena but markedly different from the behavior observed in passive glassy systems of particles~\cite{karmakar2009}. As a further check, we have plotted $\chi_4^{\text{peak}}(N,v_0)/\chi_4^{\text{peak}}(N\rightarrow\infty,v_0)$ vs. $\tau(N,v_0)/\tau(N\rightarrow\infty,v_0)$ in Fig.~(\ref{finitesizescaling}h). The data points for different $v_0$ and $N$ fall on the same curve, implying that the same correlation length governs the system-size dependence of both $\cpn$ and $\tau$. These results establish the existence of a single growing length scale that describes the growth of fluctuations and relaxation in the present system.

\section{Discussion and conclusions}
\label{discussion}

The activity-mediated unjamming of a cell monolayer is critical for several biological processes, such as cancer metastasis, embryogenesis, and development. In the early stage of EMT, the junction molecules weaken, and cells become motile. We have studied this regime and shown that the nature of the unjamming transition does not change in the presence of activity: the structure-dynamics correlation holds, and the dynamics is sub-Arrhenius. We have also characterized the higher-order correlation functions with changing $p_0$. As the system matures, junctions become firm, the dynamics becomes progressively sluggish, and the length scale, $\xi_d$, and the DH volume, $\cpn$, increase. Our results rationalize the existing experiments on confluent cell monolayers \cite{park2015}. We have also shown that a growing length scale governs the growth of DH and the relaxation time and have obtained the exponents that characterize the divergence of these quantities.

The structure-dynamics correlation has crucial significance. Many past works have demonstrated a remarkable correlation of cell shape with cellular functions, such as division plane orientation \cite{wyatt2015,hart2017,bosveld2016}, cell growth or apoptosis \cite{chen1997}, stem cell lineage \cite{mcbeath2004,wang2011}, and differentiation \cite{watt1988,roskelley1994}. In particular, cell division and apoptosis will affect the monolayer dynamics by cutting off the relaxation time \cite{ranft2010,czajkowski2019,henkes2017}. Thus, the structure-dynamics correlation of confluent monolayers is consistent with these findings. Yet, cancer cells exhibit an intricate property: they avoid the inherent defense mechanism of cell extrusion and apoptosis and continue to divide \cite{cai2021,wirtz2011}. Does this have a structural signature at the level of cell shape? Cell extrusion requires the organization of the forces and displacements. Therefore, comparing cell shapes between extruding and non-extruding cancerous cells can be instructive. Our work demonstrating the survival of the quantitative nature of the structure-dynamics correlation in the presence of activity is a first step in this direction.

Beyond the biological relevance, confluent models are also intriguing due to their fascinating physics properties. \r{In particulate systems, the experimental and simulation data follow the MCT power-law prediction for a small range of parameters and then deviate in the deeply supercooled regime. We do not see any such deviation within our simulation range, though we expect that the MCT transition will eventually be avoided even for the confluent systems. In the regime of our simulations, the glassy dynamics of the confluent systems} seems better suited for MCT \cite{ruscher2021,pandey2023}. We have shown that $\tau$ follows the power-law prediction with the same exponent as in equilibrium systems and $\cpn$ also diverges as a power law at the same critical point. Furthermore, we present the first computation of the dynamic length scale, $\xi_d$, in confluent systems via a finite-size scaling analysis \cite{karmakar2009}. We find that $\xi_d\sim \tau^{1/3}$; this is consistent with the critical nature of MCT, where the diverging time scale accompanies the divergence of a length scale. Interestingly, our results show that $(2-\eta)=1/2$; this is much smaller than the values obtained in both MCT and simulations of passive particulate systems. We have also computed a static length, $\xi_s$, and found that $\xi_d$ is proportional to $\xi_s$. This result suggests that, unlike particulate systems \cite{karmakar2009}, a unique length scale governs the dynamics.

\subsection*{Acknowledgments}
We thank Smarajit Karmakar and Thomas Voigtmann for discussions. We acknowledge the support of the Department of Atomic Energy, Government of India, under Project Identification No. RTI 4007. SKN thanks SERB for grant via SRG/2021/002014.

\appendix

\begin{figure}
	\centering
	\includegraphics[width=0.5\textwidth]{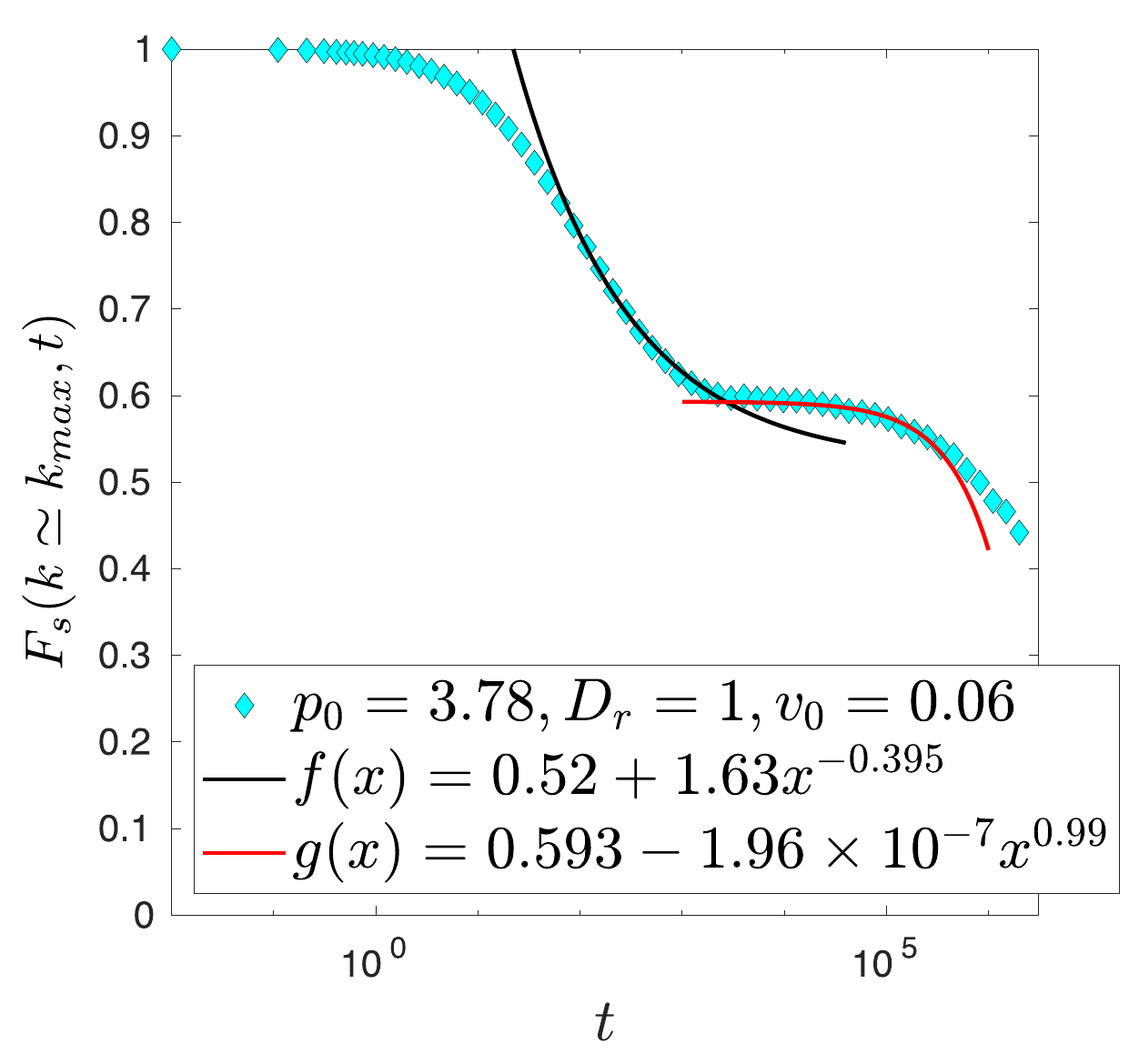}
	\caption{Test of the MCT prediction for the $\beta$-relaxation of $F_s(k,t)$ \r{for $k = 4.5$, which is close the value of $k_\text{max}$, the wavevector corresponding to the first peak of the static structure factor}. Symbols are the plot of $F_s(k,t)$ for $p_0 = 3.78$, $v_0 = 0.06$ and $D_r = 1$. We have separately fitted the early and late $\beta$-regimes with Eq. (\ref{betaregime}) (black and red lines). We obtain the parameters as follows: $f = 0.52$, $A = 1.63$, and $a = 0.395$ for the early $\beta$-regime and $f = 0.593$, $B = 1.96\times 10^{-7}$, and $b = 0.99$ for the late $\beta$-regime.}
	\label{MCTfsktscaling}
\end{figure}

\section{Determination of MCT Scaling Exponents: $a$, $b$, and $\gamma$}
\label{mctdecayexpo}

As discussed in the main text, we characterized the dynamics via the two-point self-intermediate scattering function, $F_s(k,t)$,
\begin{equation}
	F_s(k,t) = \langle \tilde{F}_s(k,t) \rangle =\frac{1}{N_{\alpha}} \langle \sum_{i=1}^{N_{\alpha}} \exp[i\mathbf{k}.(\mathbf{r}_i(0) - \mathbf{r}_i(t))]\rangle.
\end{equation}
For our binary system, we could define two different $F_s(k,t)$, one for each type of particle. They contain the same information \cite{szamel2006}. Here, we present the results for only the $\alpha$-particles.
We have chosen $k \simeq k_{\text max}$, the wavevector corresponding to the maximum of the static structure factor. We have obtained the relaxation time $\tau$ when $F_s(k,t)$ decays to $0.3$.

MCT predicts that $\tau$ diverges as a power law with an exponent $\gamma$. For an athermal active system $T \sim v_0^2$ and the relation modifies as $\tau \propto (v_0^2 - v_c^2)^{-\gamma}$. $\gamma$ is obtained by fitting the data in experiments and simulations. MCT also predicts a power-law decay for the $\beta$-regime. The early and late $\beta$-regimes of $F_s(k,t)$ are characterized by two exponents, $a$ and $b$:
\begin{equation} \label{betaregime}
	F_s(k,t) =
	\begin{cases}
		f +  At^{-a}, \,\,\, &\text{for early $\beta$-regime}, \\
		f -  Bt^{b},  \,\,\,  &\text{for late $\beta$-regime}.
	\end{cases}
\end{equation}
where $f$, $A$, and $B$ are constants \cite{gotze1992,gotzebook}. Close to the MCT transition point, $a$ and $b$ are related as,
\begin{equation}\label{verificationGamma}
	\frac{\Gamma^2(1-a)}{\Gamma(1-2a)} = \frac{\Gamma^2(1+b)}{\Gamma(1+2b)},
\end{equation}
where $\Gamma$ is the Gamma function. Furthermore, MCT also predicts a relation between $a$, $b$, and $\gamma$:
\begin{equation}\label{gammafromab}
	\gamma = \frac{1}{2a} + \frac{1}{2b}.
\end{equation}

We have tested this prediction of MCT. Figure (\ref{MCTfsktscaling}) shows the fit with the data for $p_0=3.78$ and $v_0=0.06$. We find that the values of $a$ and $b$ remain nearly constant: $a = 0.395$ and $b = 0.99$. Using these values, we find that Eq. (\ref{verificationGamma}) is valid up to the second order. In addition, we obtain $\gamma = 1.77$ from Eq. (\ref{gammafromab}). Note that this is very close to the lowest values of $\gamma$ that MCT predicts and higher than that obtained from the relaxation time data.

%

\end{document}